\documentclass[structabstract]{aa}  
\usepackage{graphicx}
\usepackage{txfonts}

\begin{document}

   \title{Electromagnetic priors for black hole spindown in searches for gravitational-waves from supernovae and long GRBs}
   
   \author{M.H.P.M. van Putten,\inst{1} M. Della Valle\inst{2,3}\and A. Levinson\inst{4}}
   \institute{Korea Institute for Advanced Study, 85 Hoegiro, Dongdaemun-gu, Seoul, 130-722 Korea, {mvputten@kias.re.kr}
   \and Istituto Nazionale di AstrofisicaÑOsservatorio Astronomico di Capodimonte, Salita Moiariello 16, I-80131 Napoli, Italy 
   \and International Center for Relativistic Astrophysics, Piazzale della Repubblica 2, I-65122, Pescara, Italy, {dellavalle@na.astro.it}
   \and School of Physics and Astronomy, Tel Aviv University, 69978 Tel Aviv, Israel, levinson@wise.tau.ac.il}
   \date{}

\abstract
{Some core-collapse supernovae appear to be hyper-energetic, and a subset of these are aspherical and associated with long GRBs.} 
{We use observations of electromagnetic emission from core-collapse supernovae and GRBs to impose constraints on their free energy source as a prior to searches for their gravitational wave emission.}
{We review these events based on a finite efficiency for the conversion of spin energy to magnetic winds powering supernovae.}
{We find that some of the hyper-energetic events cannot be powered by the spindown of rapidly rotating proto-neutron stars by virtue of their limited rotational energy. They can, instead, be produced by the spindown of black holes providing a distinct prospect for gravitational-wave emission of interest to LIGO, Virgo, and the LCGT. }
{}

   \keywords{gravitational waves -- black holes -- non-thermal emissions} 
     
   \maketitle

\section{Introduction}

The massive energy release involved in the core-collapse of a massive star indicates that there is substantional emission other than by photons of the electromagnetic spectrum as demonstrated by the $>10$ s burst in $>10$ MeV neutrinos from the supernova (SN) 1987A (\cite{bur87}). The most extreme core-collapse supernovae (CC-SNe) are believed to power long gamma-ray bursts (GRBs), which may represent the tip of the iceberg of a broader class of relativistic SNe (\cite{del10}) for which we observe a mildly relativistic aspherical ejection of the remnant stellar envelope. They may result from the spindown of a compact object through magnetic winds (\cite{bis70,leb70,ost71}), notably a (proto-)neutron star (PNS) or a magnetar, or from an accretion disk surrounding a black hole. Since the linear size of these inner engines is similar to their Schwarzschild radius, they likely emit gravitational waves.

The sensitivity of the advanced gravitational-wave detectors LIGO (\cite{bar99}), Virgo (\cite{arc04}), and the LCGT (\cite{kur10}) is expected to improve to a dimensionless strain noise amplitude $h\simeq 2\times 10^{-24}$Hz$^{-1/2}$ around 1 kHz. We will then be able to probe the local Universe within 100 Mpc for energetic bursts reaching at least a few percent of $M_\odot c^2$ in gravitational waves of frequency 50-1500 Hz. As optical-radio SN surveys improve in yield, we also expect to efficiently harvest nearby galaxies such as M51 for frequent CC-SNe, and search for their gravitational wave emission. These observations promise to help identify their mysterious explosion mechanism, the branching ratio of CC-SNe producing neutron stars and black holes, and their correlation to the supernova morphology (spherical versus aspherical), radio emission (radio-quiet versus radio-loud), and energies (non-relativistic to relativistic). 

It is found that CC-SNe produce either neutron stars or black holes with markedly different emission in gravitational waves. We recall the {\em upper bound} of $1.5-2M_\odot$ to the mass of the former (\cite{nic05,dem10}) and the {\em lower bound} $4\pm1M_\odot$ to the mass of the latter inferred from GRO J0422+32 (\cite{gel03}). The emission in gravitational-waves from either the resultant spindown of PNS or the multipole mass-moments in an accretion disk or torus surrounding a central black hole have distinct frequencies, chirps, and energy output. Any priors on the CC-SNe from the electromagnetic spectrum, including statistics on the event rates, that indicates the formation of a neutron star or a black hole is therefore a valuable asset in searches for their gravitational-wave emission by the aforementioned ground-based detectors. 

Relatively massive progenitors should produce black holes. The mass threshold is presently not tightly constrained, but given the lack of evidence of a neutron star remnant and relativistic jets in SN1987A (\cite{nis99}), it may be as low as 18$M_\odot$, which was the mass of the progenitor Sanduleak -69$^o$ 202a of SN1987A (\cite{gil87,tri88}),
although a firm lower limit of $8\pm1M_\odot$ has been set from by direct detections of red supergiant progenitors of II-Plateau SNe (\cite{sma09}). With SN1994I, SN2005cs, and SN2011dh, M51 is a nearby galaxy ($D=$8Mpc) with frequent CC-SNe. Their progenitor masses were, respectively, 13-20$M_\odot$ (\cite{you95}), $18.2\pm 1M_\odot$ (\cite{utr08}; but see \cite{tak06}), and $13\pm 3M_\odot$ (\cite{mau11}), so that they may have produced neutron stars and black holes. 

In this {\em Letter}, we focus on extracting priors to searches for gravitational waves from CC-SNe and long GRBs, using electromagnetic data to constrain the bulk kinetic energy $E_{SN}$ in SNe and the total energy $E_{tot}$ in GRB-afterglow emissions. For reference, we define $E_c=3\times 10^{52}\mbox{~erg}$ to be the maximum of the rotational energy $E_{rot}$ of a PNS. $E_c$ refers to a PNS mass of $M_\odot=1.45M_\odot$ and a radius $R=12$ km. Following \cite{hae09}, a firm upper limit of $E_{rot}=2.5 E_c$ is defined by a supermassive PNS (sPNS) of $M=2.5 M_\odot$ with $R=13$ km.
The corresponding spindown power $L_c$ (e.g. \cite{kal11}) and spindown time $T_c$ satisfy the relations $L_c\simeq 10^{51}B_{16}^2 \left({M}/{1.45M_\odot}\right)^2~\mbox{erg~s}^{-1}$ and $T_c\simeq E_c/L_c= 30B_{16}^{-2}~\mbox{s}$ for a magnetic field strength $B=B_{16}10^{16}$ erg. $E_c$ provides a dividing line between PNS ($E_{rot}\le E_c$) and rapidly rotating stellar-mass black holes ($E_{rot}=6\times 10^{54} M_1$ erg for a $M=M_110M_\odot$ extremal Kerr black hole).  Our objective is for electromagnetic observations to constrain the likelihood that supernovae and long GRBs originate from the spindown of PNPs or black holes, taking into account the finite efficiency of converting rotational energy by magnetic winds into $E_{SN}$ or $E_{tot}$. 

A black hole inner engine differs from a PNS in terms of its rotational energy and its ability to naturally produce a two-component outflow consisting of a baryon-poor jet (BPJ) emanating from the event horizon by frame dragging surrounded by a baryon-rich wind expelled by an accretion disk or torus (\cite{LE93,van03}). In contrast, a PNS tends to produce magnetic winds with rather uniform baryon-loading from polar to equatorial regions, so that the degree of baryon-loading may effectively vary only as a function of time and largely so by cooling. If associated with a (successful) GRB, the required super-strong magnetic fields tend to give rise to spindown times similar to cooling times, potentially leaving a spin energy appreciably less than $E_c$ for producing a BPJ after cooling. This questions whether PNS are able to produce GRB-SNe.

If a PNS produces a powerful SN with $E_{SN}<E_c$, one is tempted to speculate it might equally produce a low-luminosity GRB unless collapse continues to form a black hole (\cite{des08}). However, in attributing low-luminosity GRBs to inner engines with lifetimes shorter than the break-out time on the order of ten seconds (\cite{bro11}), which is similar to the cooling time of the PNS, the required magnetic field strengths exceed $10^{16}$ G. They require an amplification mechanism that is more efficient than provided by convection (\cite{des07}). Alternatively, the inner engine might be short-lived owing to a large kick velocity in the PNS or rapid spindown by means of gravitational-wave emission, or the compact object is born with relatively moderate spin. These alternatives probably cannot be verified using electromagnetic observations alone. 

While all short GRBs originate from mergers, the converse need not hold. In particular, a supernova and its non-relativistic ejecta may not be prerequisites for a successful GRB, as some of the long GRBs appear to occur with no supernova signature (\cite{del06,fyn06,gal06}). This implies that a common inner engine exists in a diversity of possible astronomical progenitors, which may include the core-collapse of massive stars and mergers alike involving rapidly spinning BHs (\cite{van08a,cai09}) and much less likely neutron stars formed in accretion-induced collapse (AIC) owing to excessive baryon-loading (\cite{des07}). In these cases, electromagnetic constraints are therefore limited to $E_{tot}$. 

The above suggests that, quite generally, the spindown of rapidly rotating black holes in tens of seconds all produce GRBs, some of which result in SNe, while the same of rapidly rotating PNS may all produce SNe and some low-luminosity GRBs. In \S2, we discuss the associated efficiencies when constraining $E_{rot}$ based on both $E_{SN}$ and $E_{tot}$. In \S3, we exemplify the use of priors in the context of gravitational radiation from black hole spindown. In \S4, we summarize our findings.

\section{Bounds on the rotational energy} 

For rotationally powered inner engines, the ejection of matter from the progenitor star is believed to be powered by magnetic outflows (instead of neutrino-driven winds) as envisioned by \cite{bis70}. As the wind propagates through the stellar envelope, it accumulates mass and decelerates.  In some cases, it may fail to break out of the envelope, but its energy, $E_w$, will eventually be released in the form of hydrodynamic ejecta that drives a supernova flash. Their momentum may be expressed by a fraction $f$ of $E_w/c$, where $f=1$ for an ultra-relativistic wind and $f=2/\beta_w$ for a nonrelativistic wind expanding at a velocity $\beta_w c$.  For nonrelativistic ejecta, momentum conservation implies that $f E_w/c=M_{ej}v_{ej}$, in terms of the mass $M_{ej}$ and velocity $v_{ej}$ of the ejecta. This in turn implies that $E_w=2E_k/(f\beta_{ej})$, where we define $E_k=M_{ej}v_{ej}^2/2$ to be the bulk kinetic energy of the ejecta, and $\beta_{ej}=v_{ej}/c$. The ratio $\eta=E_k/E_w=\frac{1}{2}f\beta_{ej}$ satisfies 
\begin{eqnarray}
\frac{1}{2}\beta_{ej} <  \eta < 1.
\label{EQN_eta}
\end{eqnarray}

For SNe powered by the spindown of a rapidly rotating PNS, we have
\begin{eqnarray}
E_w=\eta^{-1}E_{SN}\le E_c, 
\label{EQN_C1}
\end{eqnarray}
where $E_{SN}$ denotes the kinetic energy of the SN inferred from the measured values $(\beta_{ej},M_{ej})$ (i.e., $E_{SN}=E_k$). Any violation of Eq. (\ref{EQN_C1}) would indicate that a rotating BH rather than a PNS were the power source of the observed SN.  For GRB-SNe, the choice of $\eta\simeq \beta_{ej}/2$ may be justified under the assumption that the ultra-relativistic baryon-poor jet producing the prompt gamma-ray emission also powers the accompanying SN. 

The statistics of $v_{ej}$ provides a useful constraint on $\eta$. A sample of 56 core-collapse events studies by \cite{mau11} was found to have a value of $\beta_{ej}\sim 4.5\%$ for typical events, and significantly higher for very energetic events (Fig. \ref{FIG_vej}). These estimates are, however, somewhat model-dependent. 

If a multiflow structure is also envisioned for PNS, with a fast wind producing the GRB and a slower wind driving the SN, then a larger value of $\eta$ should be invoked.  In Table 1, we adopted a conservative choice (favoring the likelihood of a PNS) of $\eta=1$ for SNe without a GRB and $\eta=0.25$ ($\beta_{ej}=0.5$) for SNe associated with a GRB.

For the GRB-afterglow emission, we define the total energy as
\begin{eqnarray}
E_{tot}=E_\gamma + E_k,
\label{EQN_Et}
\end{eqnarray}
which represents the sum of the true gamma-ray and kinetic energies. In the absence of a supernova signature, the kinetic energy $E_k$ can sometimes be inferred from calorimetry derived for lower energy afterglows.
However, this only accounts for the kinetic energy of the jet producing the GRB, and may represent a lower limit in cases where a baryon-rich wind originates from a disk or torus. Those winds may be the source of the accompanying SN in GRB-SNe associations. In Table 1, we used $E_{tot}$, which was inferred from observations, as a conservative lower bound to $E_{rot}$, viz., $E_{rot}=E_{tot}$, in GRBs without a SN. The listed SNe are observationally limited to $z\le 0.17$, where a genuine absence of a supernova can be ascertained, as in, e.g., the {\em Swift} event GRB060614 at $z=0.125$ (\cite{del06}).
In contrast, low-luminosity GRBs ($E_\gamma<10^{49}$) may be attributed to baryon-rich winds (\cite{des07,des08}) or short-lived BPJ (\cite{bro11,nak11}), where the former ($\eta=1$) defines a conservative lower bound on $E_{rot}$. 
 \onlfig{1}{
\begin{figure}
\resizebox{\hsize}{!}{\includegraphics{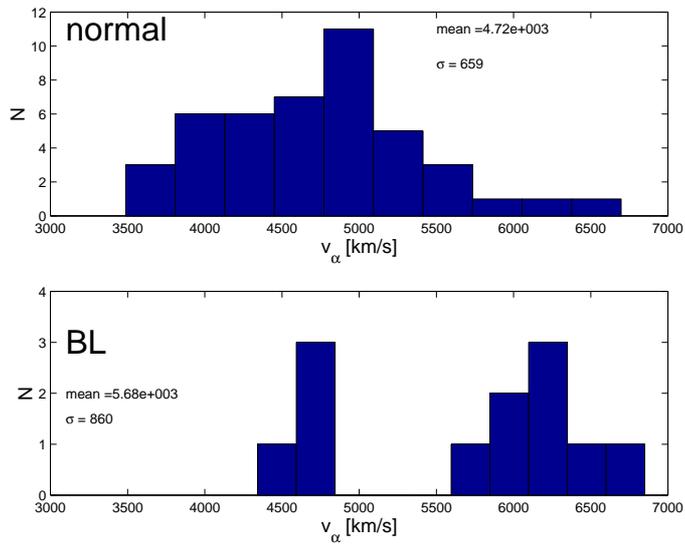}}
\caption{Histograms of the ejection velocities in core-collapse events compiled by \cite{mau10}, in 44 normal and 12 broad-line (BL) events. The mean velocity 5680 km/s of the latter is 3.9 standard deviations away from the mean velocity 4720 km/s of the first, which is indicative of a separate group of relatively hyper-energetic events.}
\label{FIG_vej}
\end{figure}
}
\begin{table*}
\caption{
References refer to SNe except for GRB 070125. Energies are in units of $10^{51}$ erg. 
\vskip-0.1in}
\centerline{
\begin{tabular}{lrrrrrrrrr}
\hline
GRB & Supernova				& Redshift $z$ & $E_\gamma$ & $E_{tot}$ & $E_{SN}$ & $\eta$ & $E_{rot}/E_c$ & Prior & Ref.\\
\hline
				& SN2005ap	& 0.283	 	& 			& 			& $>10$ 	&1	& $>0.3$	& indet & 1\\ 
       				& SN2007bi	& 0.1279 		& 			& 			& $>10$ 	&1	& $>0.3$ & indet & 1\\  
GRB 980425		&Sn1998bw 	& 0.008 	        	& $<0.001$     	& 			& 50		& 1	& 1.7		& BH & 2\\
GRB 031203		&SN2003lw 	& 0.1055	        	& $<0.17$ 	& 			& 60 		&0.25& 10	& BH & 3\\
GRB 060218 		&SN2006aj	& 0.033    	 	&$<0.04$ 		& 			& 2		&0.25& 0.25	& indet & 4 \\
GRB 100316D		&SN2006aj	& 0.0591		& 0.037-0.06	& 			& 10 		&0.25& 1.3	& BH & 5\\
GRB 030329		&SN2003dh	& 0.1685		& 0.07-0.46      & 			& 40		&0.25& 5.3	& BH & 6 \\
GRB 050820A		&			& 1.71		&  			& 42			&		&	& 1.4		& BH & 7\\ 
GRB 050904  		&			& 6.295	 	& 			&12.9		&		&	& 0.43	& indet & 7\\
GRB 070125  		&			& 1.55   		& 			& 25.3	  	& 		&	& 0.84	& indet & 7\\
GRB 080319B		&			& 0.937		& 			& 30			&		&	& 1.0 	& BH & 7\\ 
GRB 080916C		&			& 4.25		& 			& 10.2		&  		&	& 0.34	& indet & 7\\
GRB 090926A		&			& 2.1062		&       		& 14.5 		& 		&	& 0.48	& indet & 8\\
GRB 070125  		& (halo event)	& 1.55   		& 			& 25.3 		& 	 	&	& 0.84	& indet & 9\\
\hline
\hline
\end{tabular}
\label{TABLE_1}
}
\mbox{}\\\hskip0.08in
1. \cite{gal09,qui09}; 2. \cite{gal98}; 3. \cite{mal04}; 4. \cite{mas06,mod06,cam06,sol06,mir06,pia06,cob06}; 5. \cite{cho10,buf11}. 
6. \cite{sta03,hjo03,mat03}; 7. \cite{cen10}; 8. \cite{deu11}; 9. \cite{cha08}.
\end{table*}
\section{Use of a positive prior on black hole spindown}

Determining that a transient cannot originate from the spindown of a rapidly rotating PNS is an interesting prior for two reasons.
First, the gravitational-wave luminosity produced by the spindown of a PNS is highly uncertain, despite the plethora of emission channels for multiple mass-moments by acoustic modes, convection, differential rotation, and magnetic fields (\cite{ree74,owe98,cut02,cut02b,ara03,how04,dal09,and11,reg11}). Second, ruling out a PNS directs our attention to black holes as a probable alternative, which would involve different types of emission. The most relevant type would be the emission from multipole mass-moments in a  high-density inner disk or torus, which is produced by the spindown of a black hole and the resultant output of gravitational waves (\cite{van03}) of $E_{gw}[BH]\sim 10^{54} M_1$ erg representing a sizable fraction of the aforementioned $E_{rot}$ a rapidly rotating black hole. We note that $E_{gw}[BH]$ exceeds the estimated output of gravitational wave emission in hyper-accreting black holes (\cite{kob03a,kob03b}) with similar frequencies but distinct time-frequency trajectories.  In this process, BPJs emanate along the spin axis of a rotating black bole with an output $E_j$ accompanied by baryon-rich disk or torus winds for a duration set by the lifetime of the rapid spin of the black hole, which has a timescale of about one minute (\cite{van99}) and satisfies (\cite{van03,van11}) $E_{BPJ} \simeq 4\times 10^{52} M_1 \left({\theta_H}/{0.5}\right)^4~\mbox{erg},$ $E_{w}\ge 4\times 10^{53} M_1 ~\mbox{erg},$ where $\theta_H$ refers to the half-opening angle of an open magnetic flux-tube subtended to the black-hole event horizon in its lowest energy state. This outlook comfortably satisfies the required energies for hyper-energetic events, in both $E_{tot}$ and $E_{SN}$ (from $\frac{1}{2}\beta_{ej}E_w)$.

The anticipated negative chirps, associated with the expansion of the inner most stable circular orbit (ISCO) during relaxation of a near-extreme Kerr black hole to a slowly rotating black hole can be effectively searched for by a dedicated time sliced matched filtering (TSMF) method (\cite{van11}). Model results show that the spin down of black holes should be detectable to a distance about ten times further out than the spin down of neutron stars. Equivalently, they would be observable at similar event rates if the relative branching ratio of CC-SNe producing them is on the order 1:1000. 

\section{Conclusions and outlook}

Our above line of argumentation has shown that six (two) of the hyper-energetic GRB-supernovae and GRB-afterglow emissions listed in Table 1 cannot be produced by the spindown of a PNS (sPNS). For GRB-SNe with typical values of $E_\gamma$, we use $\eta=0.25$ ($\beta_{ej}=0.5$) to test the feasibility of spindown of a PNS. The priors indicating a BH follow for energies exceeding the limitations of (s)PNS, leaving aside challenges to their formation (e.g. \cite{heg05}). In Table 1, the remaining events are labeled ``indet," which refers to the indeterminate case of spindown of either a rapidly spinning black hole or PNS. 

Our multimessenger approach is complementary to others with similar scientific objectives focused on, e.g., long GRBs and high- energy neutrino emissions (\cite{bar11}); see also (\cite{cha10a,cha10b,pra09,vane09}). An interesting open question is how to incorporate our priors into model-independent all-sky searches for gravitational waves (\cite{aba11}).

We note that GRB-SNe are very rare phenomena compared to SN explosions without GRBs, as only 0.4-3\% of SNe-Ib/c become GRB progenitors (\cite{gue07}). This implies that less, or much less, than 1\% of all CC-SNe produce GRBs. Given these rates, one can predict, based on SN counts (\cite{cap99,man05,sma09b}), a rate of about 0.5 up to 3.6 GRB-SNe per year within a distance of 100 Mpc, which is the scale of the local Universe of interest to the planned advanced generation of gravitational wave detectors (e.g. \cite{van04,van11} for long GRBs). Even so, GRB-SNe are quite likely powered by the spindown of compact objects by their strongly aspherical explosions (e.g. \cite{hoe99,mae08,mod08,tau09}), whereby they are, as part of the class of CC-SNe, the most likely sources of extended emission of gravitational waves. Furthermore, hyper-energetic supernovae associated with some of the (weakly beamed) low-luminosity GRBs are of interest in view of their relative abundance, which is similar to or at least as numerous as the true rate of the (strongly beamed) long GRBs. In view of these observations, we focus specifically on the potential for using electromagnetic signatures from supernovae and the (weakly beamed) low-luminosity GRBs to predict a corresponding output in gravitational waves.

Observations of the nearby galaxy M51 have uncovered a prodigious production of core-collapse supernovae at what appears to be a rate of at least one CC-SNe per decade. The planned sensitivity of the advanced gravitational-wave detectors should put us in a strong position to probe events such as SN2011dh to establish the nature of their mysterious inner engine, and to determine whether it formed a PNS or a black hole. The SN2011dh event poses an interesting ambiguity, given the limited mass of $13\pm 3 M_\odot$ of its progenitor and, at the same time, an appreciable ejection velocity of $\beta \sim 10\%$ \cite{sod11}. More generally, these studies of M51 encourage us to develop a catalogue of similarly fertile (interacting) galaxies and to harvest their CC-SNe on a systematic basis. 

\begin{acknowledgements}
   The authors gratefully acknowledges constructive comments from both Adam Burrows and the referee.
\end{acknowledgements}

\end{document}